\documentclass[aps, pra, reprint, superscriptaddress, amsmath, amssymb, floatfix]{revtex4-1}

\usepackage{physics}
\usepackage[normalem]{ulem}
\usepackage{graphicx}
\graphicspath{ {pdffig/}, {supfig/}, {images/}, {figs/} }
\usepackage{dcolumn}
\usepackage{bm}
\usepackage{hyperref}
\usepackage{siunitx} 
\usepackage{ulem}    
\usepackage{pifont}
\usepackage{xcolor}
\hypersetup{
	colorlinks,
	linkcolor={blue!90!black},
	citecolor={blue!90!black},
	urlcolor={blue!90!black}  
}
\usepackage[caption=false]{subfig}
\makeatletter
\newcommand{\colorcaption}[2][]{%
  \begingroup%
  \renewcommand{\@caption@fignum@sep}{ (color online). }%
  \caption[#1]{#2}%
  \endgroup%
}
\makeatother

\def\Rb{$^{87}$Rb }

\def\NaRb{$^{23}$Na$^{87}$Rb }

\begin{document}

\title{Anisotropic Polarizability of Ultracold Ground-state $^{23}$Na$^{87}$Rb Molecules}

\author{Junyu Lin}
\email{jylin@phy.cuhk.edu.hk}
\affiliation{Department of Physics, The Chinese University of Hong Kong, Shatin, Hong Kong, China}
\author{Junyu He}
\affiliation{Department of Physics, The Chinese University of Hong Kong, Shatin, Hong Kong, China}
\author{Xin Ye}
\affiliation{Department of Physics, The Chinese University of Hong Kong, Shatin, Hong Kong, China}
\author{Dajun Wang}
\email{djwang@cuhk.edu.hk}
\affiliation{Department of Physics, The Chinese University of Hong Kong, Shatin, Hong Kong, China}
\affiliation{The Chinese University of Hong Kong Shenzhen Research Institute, Shenzhen, China}

\date{\today}

\begin{abstract}
We report measurements of the ac polarizabilities of ultracold ground-state $^{23}\rm{Na}^{87}\rm{Rb}$ molecules. While the polarizability of the ground rotational state $J = 0$ is isotropic, that of the first excited rotational state $J = 1$ is anisotropic and depends strongly on the light polarization angle. We obtain both polarizabilities precisely by combining trap oscillation frequency measurement and high resolution rotational spectroscopy driven by microwave. With the optimized light polarization angle and intensity combination, the nonuniformity of the differential ac Stark shift between the two rotational states is minimized and the rotational coherence time is observed to be the longest.     
\end{abstract}

\maketitle

\section{Introduction}
In recent years, ultracold polar molecules (UPMs) have been attracting more and more attentions. The strong, long-range, and anisotropic electric dipole-dipole interaction between UPMs is a valuable resource for applications in quantum simulation~\cite{QuantumMagnetism,micheli2006toolbox,Topologicalphasesinultracoldpolar,RealizingFractionalChern} and quantum computation~\cite{QuantumComputationwithPolar,Schemesforrobustcom,zhu2013implementation,herrera2014infrared,ni2018dipolar}. In addition, UPMs are also a nature candidate for  ultracold chemistry~\cite{ospelkaus2010quantum,ye2018collisions,hu2019direct,LossofUltracoldRbCs} and precision measurement~\cite{PrecisionTestofMassRatio,hudson2011improved}. While most of the studies so far are theoretical, with more and more UPM species become available in the last several years following the exciting developments in both association of ultracold atoms~\cite{KRb2008,RbCs201411,RbCs201412,NaK201505,NaRb201605,NaLi201710,NaK201801,NaK2020} and direct laser cooling of molecules~\cite{DirectSrF,DirectCaF,HighDensityCaF,3DMOTYO}, some important characteristics of this new form of matter are starting to be revealed experimentally. 

Many of the proposed applications of UPMs also rely on manipulation of their rich internal degrees of freedom. One of the most important internal structures of polar molecules is their rotational levels~\cite{internalKRb,internalNaK,NaRbinternalstate} which can be readily coupled with a microwave (MW) field for inducing the dipole-dipole interaction~\cite{spinexchangeKRb} and for realizing various topological phases~\cite{Topologicalphasesinultracoldpolar,RealizingFractionalChern}. To explore these ideas, however, it is often necessary to maintain a long coherence time between the rotational levels. For optically trapped UPMs, a main rotational decoherence mechanism with MW coupling is the spatially-dependent rotational transition frequencies caused by the non-uniform trapping light. This issue stems from the polarizabilities of the ground and excited rotational levels which, in general, are not the same. To suppress this one-body decoherence, a common method is to make use of the anisotropic nature of the polarizability of the rotational excited state. By setting the angle between the trap laser polarization and the quantization axis defined by the magnetic field to the ``magic'' angle~\cite{MagicAngleTheory}, the ac polarizabilities between the excited and ground rotational state can be tuned to be the same. In several UPM species, this method has been proven to be able to extend the rotational coherence time significantly~\cite{KRbMagicAngle,RbCsacStark2017,RbCsacStark2020}.

In this work, we investigate the polarizabilities of ground-state $^{23}\rm{Na}^{87}\rm{Rb}$ molecules at the optical trap wavelength of $\lambda = {1064.17}\mathrm{~nm}$. The isotropic polarizability of the $J = 0$ rotational ground state is extracted from the harmonic trap frequency, while the anisotropic polarizability of the $J = 1$ rotational level at different intensities and polarization angles are obtained with the help of MW spectroscopy. With these results, we study the rotational coherence and observe that matching the polarization angle with the trap intensity is important to obtain the longest coherence time. This result is explained well by the nonlinear intensity dependence of the ac Stark shift in the $J = 1$ level. 



The rest of this paper is organized as follows. In Section \ref{section:acstark}, a Hamiltonian describes the hyperfine and rotational structure of $^{23}\rm{Na}^{87}\rm{Rb}$ molecules in magnetic and optical fields is presented. In Section \ref{section:pol}, our measurements of isotropic and anisotropic polarizabilities are reported. In Section \ref{section:coherence}, we present the study on optimizing the rotational coherence. Finally, we conclude this work in Section \ref{section:conclusion}.

\section{molecular Hamiltonian and ac stark effect}
\label{section:acstark}

The Hamiltonian for an optically trapped vibronic ground-state NaRb molecule in magnetic field can be written as
\begin{equation}
H=H_{rot}+H_{hf}+H_{Zeeman}+H_{ac}.
\label{eq:Hfull}
\end{equation}
Here $H_{rot}$ and $H_{hf}$ are the rotational and hyperfine structures, $H_{Zeeman}$ is from the Zeeman effect, and $H_{ac}$ is due to the ac Stark effect.

The rotational Hamiltonian is written as $H_{rot}=B_{v}J(J+1)$, with $B_{v}$ the rotational constant. In this work, we focus on two rotational subspaces, with quantum number $J=0, m_J=0$ and $J=1, m_J=0, \pm{1}$. Here $m_J$ is the projection of the rotational quantum number $J$. For the NaRb molecule, $B_{v}$ is about 2.090 GHz and the rotational transition frequency between $J = 0$ and $J = 1$ is $2B_{v} = 4.179$~GHz~\cite{NaRbinternalstate}.

The hyperfine Hamiltonian $H_{hf}$ contains several terms including the electric quadrupole interaction, the spin-rotation coupling, and the tensor and the scalar coupling between the two nuclei~\cite{NaRbinternalstate}. The electric quadrupole interaction term is characterized by the coupling constants $(eqQ)_{Na}$ and $(eqQ)_{Rb}$ for $^{23}\rm{Na}$ and $^{87}\rm{Rb}$ nucleus, respectively. This term dominates $H_{hf}$ for $J = 1$ but vanishes for $J = 0$, while the other terms are much smaller. The nuclear spins of both $^{23}\rm{Na}$ and $^{87}\rm{Rb}$ are $I=3/2$. Every eigenstate of the full Hamiltonian can be represented as a linear combination of the bare states $\ket{J,m_{J},m_{I}^{Na},m_{I}^{Rb}}$, with $m_{I}^{Na}$ and $m_{I}^{Rb}$ the projections of the nuclear spins. In this work, only hyperfine levels with dominating $m_{I}^{Na} = m_{I}^{Rb} = 3/2$ components are used and they will be simply denoted by $\ket{J, m_J}$.

In the $J = 0$ rotational subspace, the ac Stark Hamiltonian for $\ket{0,0}$ state versus the trap laser intensity $I$, denoted as $H^{(0)}$, is given by 
\begin{equation}
H^{(0)} = -\alpha_{0}I
\end{equation}
where $\alpha_{0} = ({\alpha_{||}}+{2\alpha_{\perp}})/{3}$ is the isotropic polarizability. Here, $\alpha_{||}$ and $\alpha_{\perp}$ are respectively the polarizabilities parallel and perpendicular to the intermolecular axis, with contributions from all relevant excited states. We note that although $\alpha_{0}$ is a constant for $\ket{0,0}$, with the non-uniform $I$ of the optical trap, the ac Stark shifts of the trapped sample are also spatial dependent.

\begin{figure}[t]
    \centering
    \includegraphics[width=0.9\linewidth]{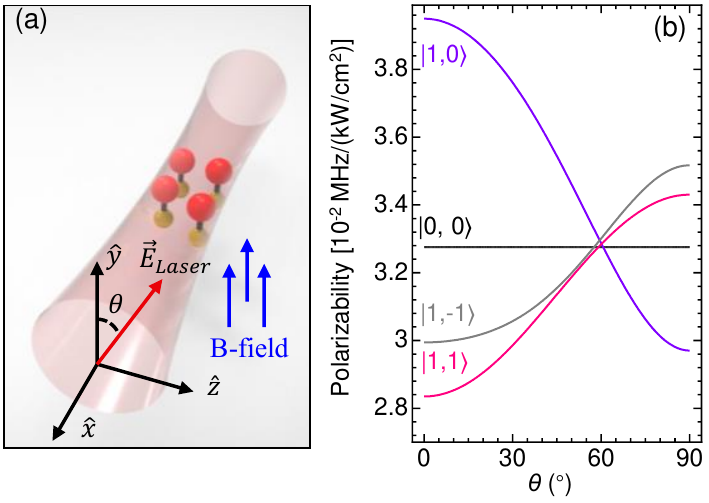}%
\caption{Anisotropic polarizability and ``magic'' angle for ultracold \NaRb molecules. 
    (a) Schematic experimental setup. The ground-state \NaRb molecules are confined in an optical dipole trap formed by a focused laser beam propagating along the $\hat{x}$-direction. The angle $\theta$ between the laser polarization $\vec{E}_{Laser}$ and the quantization axis defined by the magnetic field in $\hat{y}$-axis can be adjusted by a half-waveplate. (b) Polarizabilities for \NaRb molecules in $\ket{0, 0}$, $\ket{1,0}$, and $\ket{1,\pm{1}}$ states versus $\theta$ at a uniform intensity of ${5}\mathrm{~kW/cm^2}$. The $J = 0$ and $J = 1$ polarizabilities intersect each other at the ``magic'' angles.   
    }
\label{fig:intro}
\end{figure}

For the $J = 1$ levels, by choosing bare states with $m_J=0$, $1$ and $-1$ as the basis, the ac Stark Hamiltonian is
\begin{equation}
H_{ij}^{(1)} = -\alpha_{0}I\delta_{ij} -\alpha_{1}I d_{ij},
\end{equation}
with $\alpha_{1} = 2(\alpha_{||}-\alpha_{\perp})/3$ the anisotropic polarizability. The anisotropy of the second term is manifested in the matrix elements $d_{ij} = d_{ji}$ with
\begin{align}
&d_{11}= \frac{3\cos^{2}\theta-1}{5}, &\quad \nonumber d_{22}&=d_{33}=-\frac{3\cos^{2}\theta-1}{10}, \nonumber\\
&d_{23}=-\frac{3\sin^{2}\theta}{10}, &\quad \nonumber d_{12}&=-d_{13}=-\frac{3\sqrt{2}\sin\theta\cos\theta}{10}. \\
\end{align}
which depend on the angle $\theta$ [Fig.~\ref{fig:intro}(a)] between the polarization of the trapping light and the magnetic field. 

At the typical light intensities of several $\mathrm{kW/cm^{2}}$ used in the experiment, the off-diagonal elements of $H^{(1)}$ become comparable to the hyperfine splitting of the $J = 1$ level and, more importantly, they introduce significant couplings between the hyperfine levels in addition to those caused by $H_{hf}$~\cite{RbCsacStark2017,NaKMaictrap}.  As a result, the ac Stark shift $E_{\ket{1,m_J}}$ of the $J = 1$ molecule becomes nonlinear and the differential ac Stark shifts between $\ket{0,0}$ and $\ket{J,m_J}$ can be expressed as,
\begin{equation}
    \Delta E
    = \Delta \alpha I
    +\beta_{\ket{1,m_J}} I^2
    + \mathcal{O}(I^3).
\label{eq:dV}
\end{equation}
Here $\Delta \alpha$ is the difference between the (scalar) polarizability of $\ket{1,m_J}$ and $\alpha_0$, the polarizability of $\ket{0,0}$. The second-order term coefficient $\beta_{\ket{1,m_J}}$ is the hyperpolarizability for the $\ket{1,m_J}$ state. For the $\ket{0,0}$ state, the hyperpolarizability is zero.

For a uniform intensity, by adjusting the light polarization angle $\theta$, it is possible to find a ``magic'' angle to eliminate $\Delta E$. Considering the dominating linear term in Eq.~\ref{eq:dV} only, this can be achieved by making $\Delta \alpha = 0$. In this case, the ``magic'' angle should satisfy $3 \rm{cos}^2\theta - 1 = 0$ which corresponds to $\theta = 54.7^{\circ}$, regardless of the trap intensity. A more precise ``magic'' angle with the nonlinear terms taken into account should be obtained by solving the full Hamiltonian in Eq.~\ref{eq:Hfull} numerically. As shown in Fig.~\ref{fig:intro}(b), for $^{23}\rm{Na}^{87}\rm{Rb}$ trapped by a trap light intensity of 5~kW/cm$^2$, with the $\alpha_0$ and $\alpha_1$ measured in this work, the ``magic'' angle is calculated to be 61$^{\circ}$ for the $\ket{0,0}\leftrightarrow \ket{1,0}$ transition. This is very different from the value calculated with the linear term only and demonstrate the importance of including the non-linear contributions.

The most significant complication caused by the non-linear intensity dependence is the ``magic'' angle also becomes intensity dependent. As a result, in the optical trap with non-uniform intensity, $\Delta E$ cannot be totally eliminated across the whole sample. In general, this will always lead to rotational decoherence. The best strategy is to find the $\theta$ which makes the intensity-dependent ``local'' polarizability $\partial{E_{\ket{1,m_J}}}/\partial{I}$ of the $\ket{1,m_J}$ state for the light intensity at the center of the sample equals to $\alpha_0$, i.e., $\partial{\Delta E_{\ket{1,m_J}}}/\partial{I} = 0$. With this $\theta$ fixed, the variation of $\Delta E$ across the whole sample is the smallest, i.e., it is as uniform as possible; thus the rotational coherence time should be the longest.

\section{measurement of the isotropic and anisotropic polarizabilities}
\label{section:pol}


\subsection{Isotropic Polarizability $\alpha_{0}$}
\label{iso}

In the focused beam optical dipole trap, the oscillation frequency of the $\ket{0,0}$ state \NaRb is $\propto \sqrt{\alpha_0 I/m_{\rm NaRb}}$, with $m_{\rm NaRb}$ the mass of \NaRb. It is thus possible to obtain $\alpha_0$ from the trap frequency if the trap light profile and power at the position of the sample are known. To this end, we characterize the optical dipole trap profile with \Rb atoms taking the advantage of its well known ac polarizability~\cite{Safronova2004}. Equivalently, even without knowing the trap beam power, by measuring the oscillation frequencies of \Rb and \NaRb independently in the same trap configurations, $\alpha_0$ can be obtained directly.

The experimental procedure for creating the optically trapped samples of absolute ground-state $^{23}\rm{Na}^{87}\rm{Rb}$ molecules has been discussed in detail before~\cite{NaRb201605}. Starting from an ultracold mixture of $^{23}\rm{Na}$ and $^{87}\rm{Rb}$ atoms, we first create weakly-bound Feshbach molecules with magneto-association, and then transfer them into $\ket{0,0}$ state with a stimulated Raman adiabatic passage (STIRAP). The optical dipolar trap is formed by crossing two focused horizontal laser beams at $\SI{90}{\degree}$. 

\begin{figure}[t]
    \centering
    \includegraphics[width = 0.9 \linewidth]{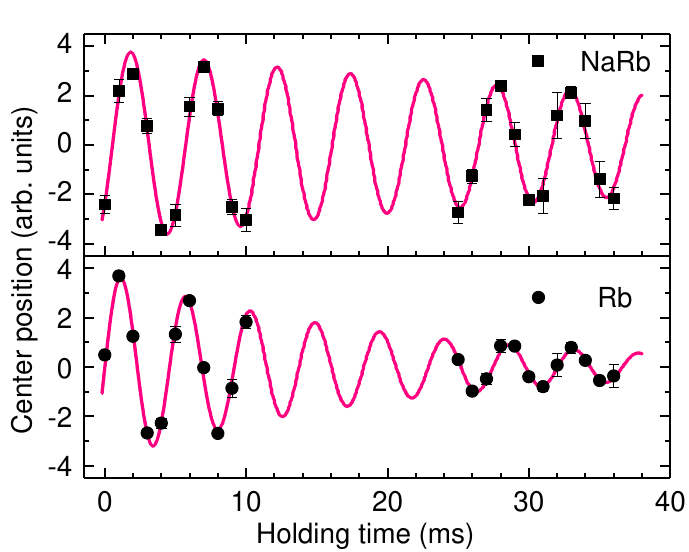}
\caption{Measuring the isotropic polarizability $\alpha_0$ of \NaRb molecules from slosh motion frequencies in the optical trap. The upper (lower) panel is the sample center position in the y direction for $\ket{0,0}$ state $^{23}\rm{Na}^{87}\rm{Rb}$ molecules ($^{87}\rm{Rb}$ atoms). Error bars represent one standard deviation for three times measurement. The solid curves are fits to the damped sine function for extracting the trap frequencies. The two measurements are performed with the same optical trap parameters. $\alpha_0$ is obtained from these trap frequencies with Eq.~\ref{eq:iso}.   
    } 
\label{fig:slosh}
\end{figure}

To measure the trap oscillation frequency, we initiate the slosh motion of the sample in the vertical direction by applying a short perturbation to the trapping light. Afterwards, we hold the sample for different times before turning off the trap and measuring the center position of the sample. For the detection, we transfer the ground-state molecules back to Feshbach molecules with a reversed STIRAP. The Feshbach molecules are then dissociated into atoms which are probed by standard absorption imaging method. The total time-of-flight between the trap turning off and the detection is 2.4~ms. The resulting slosh motion for \NaRb is shown in the upper panel of Fig.~\ref{fig:slosh}. The slosh motion for \Rb atoms in the same trap light power is obtained in a similar way and the result is shown in the lower panel of Fig.~\ref{fig:slosh}. 

Fit these slosh motions with the damped sine function, we extract the trap frequency $\omega_{\rm NaRb} = 2\pi\times 193.1(4)$~Hz for \NaRb and $\omega_{\rm Rb} = 2\pi\times 218.6(7)$~Hz for \Rb~\cite{Safronova2004}. From these results, the isotropic polarizability of \NaRb can be obtained from the relation:  
\begin{equation}
\frac{\alpha_{0}}{\alpha_{\rm{Rb}}} = \frac{m_{\rm{NaRb}} \omega_{0}^{2}}{m_{\rm{Rb}}\omega_{\rm{Rb}}^{2}}  \frac{w_{y}^{2}- \frac{4g^{2}} {\omega_{\rm{Rb}}^{4}}} {w_{y}^{2}-{\frac{4g^{2}}{\omega_{0}^{4}}}} e^{\frac{2 g^{2}}{w_{y}^{2}}(\frac{1}{\omega_{0}^{4}}-\frac{1}{\omega_{\rm{Rb}}^{4}})}.
\label{eq:iso}
\end{equation}
Here $\alpha_{\rm{Rb}} = {3.240\times10^{-2}}\mathrm{~MHz/(kW/cm^{2})} = {691.2}{\mathrm{~a.u.}}$ is the polarizability of \Rb at 1064.17~nm, $m_{\rm Rb}$ is the atomic mass of $^{87}$Rb, and $w_y = 66.2(1.3)\mathrm{~\mu m}$ is the geometric mean of the trap beam radius along the vertical direction and at the position of the sample. We also take into account the sags in the vertical direction caused by gravity which are slightly different for \Rb and \NaRb, with $g = 9.8$~m/s$^2$ the gravitational acceleration. The measured $\alpha_0$ of the $\ket{0,0}$ molecules is ${699(5)}{\rm{~a.u.}}$, which agrees with the theoretical value $\SI{674.17}{\rm{a.u.}}$ considering the accuracy of the calculation~\cite{theoryPolari}.


\subsection{Anisotropic Polarizability $\alpha_{1}$}

\begin{figure}[t]
    \centering
    \includegraphics[width=0.9\linewidth]{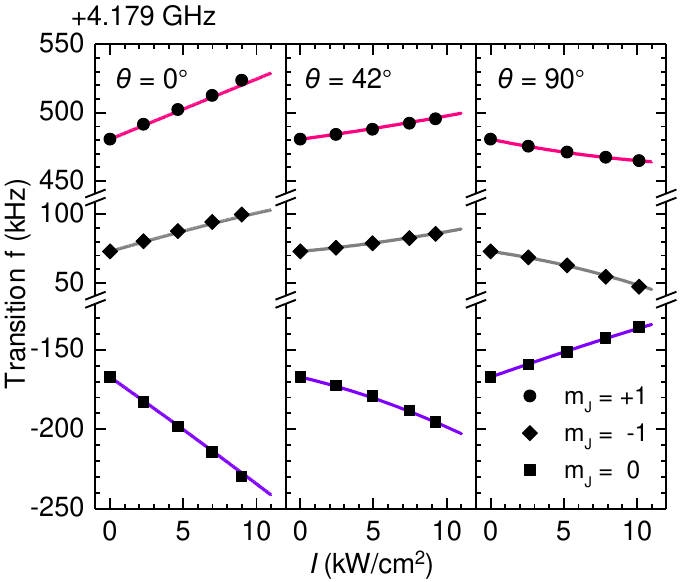}
    \caption{
    Measurement of the anisotropic polarizability from the ac Stark shifted rotational transition frequencies. The transition frequencies from $\ket{0,0}$ to $\ket{1,0}$ and $\ket{1,\pm 1}$ versus trap intensity are obtained for $\theta = \SI{0}{\degree}$ (left), $\SI{42}{\degree}$ (middle) and $\SI{90}{\degree}$ (right). The data points are extracted from fitting to the MW-driven Rabi lineshapes. The error bars are on the sub-kHz level which are smaller than the symbols in this scale. The solid curves are from numerical fitting to the full Hamiltonian for obtaining the anisotropic polarizability of the three $J = 1$ states. }
    \label{fig:MWtransition}
\end{figure}

The anisotropic polarizabilities of the three $\ket{1,m_J}$ states are obtained from measurements of their ac Stark shifts. To this end, we switch to a single beam optical trap immediately after creating the $\ket{0,0}$ state \NaRb sample. A ${200}\mathrm{~\mu s}$ MW pulse is then applied to drive $\ket{0,0} \rightarrow \ket{1,m_J}$ transitions at various light intensity and polarization angle combinations. We scan the MW frequency and fit the obtained MW Rabi lineshape with a Gaussian to extract the resonant transition frequency. Due to the typically non-uniform ac Stark shift, the Rabi lineshape is broadened. The measured transition center corresponds to the intensity at the center of the sample. Fig.~\ref{fig:MWtransition} shows the transition frequency for $\theta = \SI{0}{\degree}, \SI{42}{\degree}$ and $\SI{90}{\degree}$ for trap intensities from 0 to about ${10}\mathrm{~kW/cm^2}$. The polarization angle $\theta$ is varied by adjusting a half-waveplate in the optical trap beam path which allows us to tune $\theta$ with a $\SI{1}{\degree}$ precision. 

The zero intensity data are taken by switching off the optical trap and are thus free of ac Stark shifts. Fit this set of data with Eq.~\ref{eq:Hfull}, we obtain the the several important constants $B_{v} = {2.08966034(1)}\mathrm{~GHz}$, $(eqQ)_{Na} ={-0.15906(4)}\mathrm{~MHz}$, and $(eqQ)_{Rb}= {-3.0098(1)}\mathrm{~MHz}$. We note that these values are slightly different from the previously measured ones~\cite{NaRbinternalstate} which we attribute to the different MW Rabi frequencies used in these two measurements. For the current work, a low Rabi frequency of ${2}\mathrm{~kHz}$ is used which reduces greatly off-resonant coupling to other hyperfine levels.    

With these constants fixed, we fit the 9 curves in Fig.~\ref{fig:MWtransition} simultaneously with Eq.~\ref{eq:Hfull}. The fitting gives the anisotropic polarizability $\alpha_{1} = {2.20(3)\times10^{-2}}\mathrm{~MHz/(kW/cm^2)}$ which is equivalent to 469(7)~a.u.


\section{The ``magic'' angle and the Coherence Time}
\label{section:coherence}


\begin{figure}[h]
  \centering
  \includegraphics[width=0.9\linewidth]{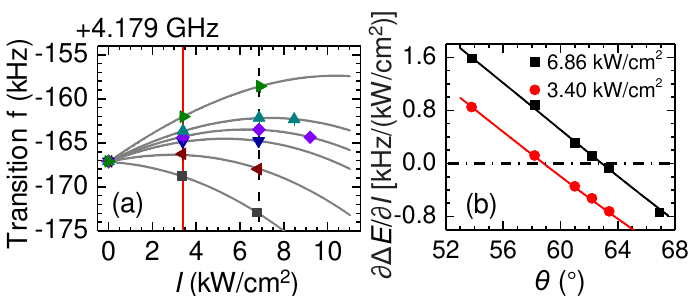}
\caption{The intensity-dependent ``magic'' angle. (a) The $\ket{0,0} \rightarrow \ket{1,0}$ transition frequency at different trap intensities for $\theta$ (from bottom to top) at 53.8$^\circ$, 58.2$^\circ$, 61.0$^\circ$, 62.2$^\circ$, 63.4$^\circ$, and 66.9$^\circ$. The solid curves are numerical solutions of the full Hamiltonian with the experimentally measured parameters. The polarization angles $\theta$ are adjusting by typically less than 1$^\circ$ in the calculation for the best agreement. The red solid and black dashed vertical lines mark intensities of 3.4~kW/cm$^2$ and 6.8~kW/cm$^2$ used in (b) which shows $\partial{\Delta E}/\partial{I}$ verse $\theta$ at these two intensities. The solid lines in (b) are from the theoretical calculation. The ``magic'' angle $\theta$ at each intensity occurs at the zero-crossing point. }
\label{fig4}
\end{figure}

\begin{figure}[h]
  \centering
  \includegraphics[width=0.9\linewidth]{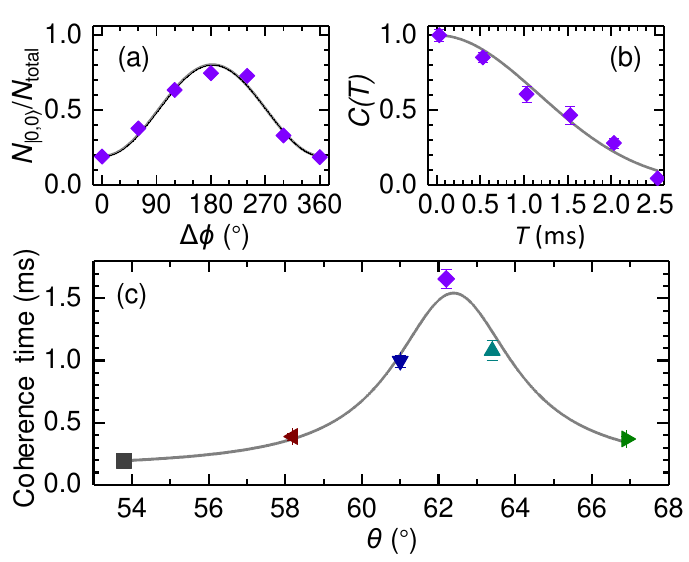}
\caption{
    Extending the rotational coherence time with the ``magic'' angle polarization. 
    (a) A Ramsey fringe at $\theta = \SI{62.2}{\degree}$ and a selected trap intensity of ${6.86}\mathrm{~kW/cm^2}$. The solid curve is the fit to Eq.~\ref{contrast} for extract the contrast $C(T)$. (b) Contrast decay of the Ramsey fringe versus $T$. The solid curve is from the Gaussian fitting for obtaining the 1/e coherence time. (c) Coherence time versus $\theta$. The maximum coherence time 1.65~ms is obtained at $\SI{62.4}{\degree}$ from the fitting with a Lorentzian (the solid curve). Error bars in (a) are one S.D., while those in (b) and (c) are from the corresponding fitting. }
\label{fig:coherence}
\end{figure}

With $\alpha_{0}$ and $\alpha_{1}$ known, we now use the ``magic'' angle idea to extend the rotational coherence of the $\ket{0,0} \leftrightarrow \ket{1,0}$ transition. In Fig.~\ref{fig4}(a), we summarize the rotational transition frequency for six $\theta$ values around 60$^\circ$. Following the discussion on the intensity-dependent ``local polarizability'' in Eq.~\ref{eq:dV}, for each $I$, the longest coherence time should occur at a specific $\theta$ satisfying $\partial{\Delta E}/\partial{I} = 0$ which corresponds to to the vertex of the measured transition frequency.

Figure~\ref{fig4}(b) shows the $\partial{\Delta E}/\partial{I}$ extracted from the data in Fig.~\ref{fig4}(a) for two different intensities. When $\theta$ is tuned from 58.2$^\circ$ to 66.9$^\circ$, a ``magic'' angle with $\partial{\Delta E}/\partial{I} = 0$ appears. For $I = 6.86~\rm{kW/cm^2}$, the ``magic'' angle is at $\theta = 62.9^\circ$. At $I = \rm{3.40~kW/cm^2}$, the ``magic'' angle is changed to $58.8^\circ$. This is thus a direct demonstration of the intensity dependence of the ``magic'' angle and the nonlinear nature of the anisotropic ac Stark shift.

To measure the rotational coherence and its dependence on $\theta$, we use a fixed intensity of $6.86(18)\mathrm{~kW/cm^2}$. For each $\theta$ here, we measure the coherence time from the contrast decay of the on-resonance MW Ramsey fringe obtained by sweeping the relative phase $\Delta\phi$ between the two $\pi/2$ pulses. As shown in Fig.~\ref{fig:coherence}(a), by scanning $\Delta\phi$ at a fixed free evolution time $T$ between the two pulses, we obtain a fringe which we fit with
\begin{equation}
    \frac{N_{\ket{0,0}}}{N_{\mathrm{total}}}
    =\frac{1}{2}\Big[1-C(T)\cos (\Delta \phi)\Big]
   \label{contrast}
\end{equation}
to get the contrast $C(T)$. Here $N_{\ket{0,0}}/N_{\rm{total}}$ is the fractional molecule number in $\ket{0,0}$. Fig.~\ref{fig:coherence}(b) shows $C(T)$ at different $T$ obtained following this procedure at $\theta = 62.2^\circ$. The coherence time at this $\theta$ is then extracted from the 1/e time constant of the Gaussian fitting to $C(T)$.


Figure \ref{fig:coherence}(c) summarizes the coherence time measurement versus $\theta$ at the selected light intensity. The maximum coherence time of 1.65(7)~ms is observed at $\theta = 62.4^\circ$ where the $\partial{\Delta E}/\partial{I} = 0$ condition is satisfied. This ``magic'' angle is in good agreement with the predicted ``magic'' angle of $62.9^\circ$ for this intensity [Fig.\ref{fig4}(b)]. For smaller $\theta$, $\partial{\Delta E}/\partial{I}$ is negative at the selected intensity, while for larger $\theta$, $\partial{\Delta E}/\partial{I}$ becomes positive. For both cases, shorter coherence times are observed as a result of the non-zero $\partial{\Delta E}/\partial{I}$.

\section{Conclusion}
\label{section:conclusion}
We studied the isotropic and anisotropic polarizabilities of \NaRb in an optical dipole trap and developed a full understanding on optimizing the rotational coherence time by considering both trap intensity and the ``magic'' polarization angle. This investigation lays the foundation for extending the rotational coherence in other forms of optical traps, such as optical tweezers and a 3D optical lattice. The long coherence time is important for investigating the physics related to the dipolar exchange interaction~\cite{spinexchangeKRb,Topologicalphasesinultracoldpolar,RealizingFractionalChern} and for realizing quantum logic gates with UPMs~\cite{ni2018dipolar,Hughes2020,Sawant2020}. With its large permanent electric dipole moment, \NaRb is a very promising candidate for realizing these exciting proposals. Currently, the coherence time is limited by the ac Stark effect induced hyperfine coupling. Applying a moderate electric field or a stronger magnetic field will help to decouple the hyperfine levels and suppress the hyperpolarizability effect. In future studies, these improvements should lead to an even longer coherence time~\cite{NaKMaictrap, RbCsacStark2020}.

\begin{acknowledgments}
This work is supported by Hong Kong RGC General Research Fund (grants 14301818, 14301119, and 14301815) and the Collaborative Research Fund C6026-16W.
\end{acknowledgments}

\appendix


%

\end{document}